# Influence of Local Defects on the Dynamics of O-H Bond Breaking and Formation on a Magnetite Surface

*Alexander Bourgund,[1] Barbara A. J. Lechner,[1,*] Matthias Meier,[2,3] Cesare Franchini,[3] Gareth S. Parkinson,[2] Ueli Heiz[1] and Friedrich Esch[1]*

[1] Chair of Physical Chemistry, Department of Chemistry & Catalysis Research Center, Technical University of Munich, 85748 Garching, Germany

[2] Institute of Applied Physics, Technische Universität Wien, 1040 Vienna, Austria

[3] Center for Computational Materials Science, Faculty of Physics, University of Vienna, 1090 Vienna, Austria

\* bajlechner@tum.de

ABSTRACT. The transport of H adatoms across oxide supports plays an important role in many catalytic reactions. We investigate the dynamics of H/Fe$_3$O$_4$(001) between 295 and 382 K. By scanning tunneling microscopy at frame rates of up to 19.6 fps, we observe the thermally activated switching of H between two O atoms on neighboring Fe rows. This switching rate changes in proximity to a defect, explained by density functional theory as a distortion in the Fe–O lattice shortening the diffusion path. Quantitative analysis yields an apparent activation barrier of $0.94 \pm 0.07$ eV on a pristine surface. The present work highlights the importance of local techniques in the study of atomic-scale dynamics at defective surfaces such as oxide supports.

INTRODUCTION

Iron oxides have attracted considerable research interest[1–7] in recent years as supports for heterogeneous catalysts due to their natural abundance[8,9] und non-toxicity. Reactions involving hydrogenation/dehydrogenation steps often require hydrogen diffusion on the support before the hydrogen evolution can take place. A fundamental understanding of the transport behavior of hydrogen on oxide surfaces can help us understand and optimize reaction mechanisms. Sophisticated integral techniques can provide some insights. In a recent study, Karim and coworkers designed a clever experimental layout to distinguish the hydrogen transport between points of different distances.[10] They found that hydrogen diffusion on the reducible titanium oxide surface is ten orders of magnitude faster than on the non-reducible aluminium oxide. In a recent temperature programmed desorption (TPD) experiment, Walenta and coworkers showed that even sparsely distributed Pt co-catalysts on a $TiO_2$(110) surface can facilitate the recombinative desorption of $H_2$ in the photocatalytic methanol reforming, thus recovering the catalytically active site and closing the catalytic cycle.[11] On some iron oxide surfaces, similar long-range diffusion of hydrogen was observed. On a thin FeO(111) film on Pt(111), for example, hydrogen diffusion occurs readily even at cryogenic temperatures.[12] Using scanning tunneling microscopy (STM), the authors of that study showed that the diffusion could further be facilitated by the presence of co-adsorbed water. On the magnetite $Fe_3O_4$(001) surface, however, only confined diffusion of hydrogen atoms was observed below the water desorption temperature.[13–15] Here, hydrogen can bind to two specific sites in the unit cell of the surface reconstruction and switch between them reversibly.[16,17] In the present work, we investigate the switching between these two adjacent sites at elevated temperatures – i.e. where catalytic reactions most often occur.

Bulk magnetite crystallizes in an inverse spinel structure where $Fe^{2+}$ occupies octahedral sites and $Fe^{3+}$ occupies tetrahedral and octahedral sites with a ratio of 1:1.[18–20] Its (001) surface reconstructs into the subsurface cation vacancy (SCV) structure, where the iron atoms of the uppermost layers are all $Fe^{3+}$, corresponding to a $(\sqrt{2} \times \sqrt{2})R45°$ reconstruction.[21] Empty-state STM images show characteristic undulating rows of the octahedral iron atoms (see Figure 1). $Fe_3O_4(001)$ has a rich defect chemistry that has been discussed in the literature at length.[15,22–24] We focus here on the so-called unreconstructed unit cells of the SCV structure, a local region on the surface where the interstitial tetrahedral Fe atom ($Fe_{tet}$) of the SCV reconstruction is missing and instead two additional Fe atoms are located in octahedral sites ($Fe_{oct}$) of the third atomic layer.[8,22] The defect is the result of a surplus Fe atom in one $Fe_{oct}$ site (labelled $+Fe_{oct}$ in Figure 1d) which causes the Fe atom in the tetrahedral site of the second layer to move to the other octahedral site in the third layer (indicated by the dashed arrow in Figure 1d). This unreconstructed defect region corresponds to a genuine $(1 \times 1)$ bulk structure unit cell and appears as a double-lobed elongated protrusion in STM (highlighted by the black oval in Figure 1a). Surface hydroxyls (OH groups), in contrast, appear as single protrusions (marked by white ovals in Figure 1a-b). These surface hydroxyl groups result from H adatoms which are formed during the crystal preparation process, where water from the residual gas atmosphere can dissociate at oxygen vacancies.[15]

Here, we present the first quantitative analysis of the H adatom switching between two adjacent O atoms in the $Fe_3O_4(001)$ unit cell to determine the energetics involved in this process. Our FastSTM measurements show that the switching process is an activated one. Furthermore, we find that at a given temperature, the switching frequency increases or decreases depending on the local chemical environment of the hydroxyl group. We therefore compare the switching behavior of hydroxyl groups on the pristine SCV $Fe_3O_4(001)$ surface with ones close to an unreconstructed unit cell.

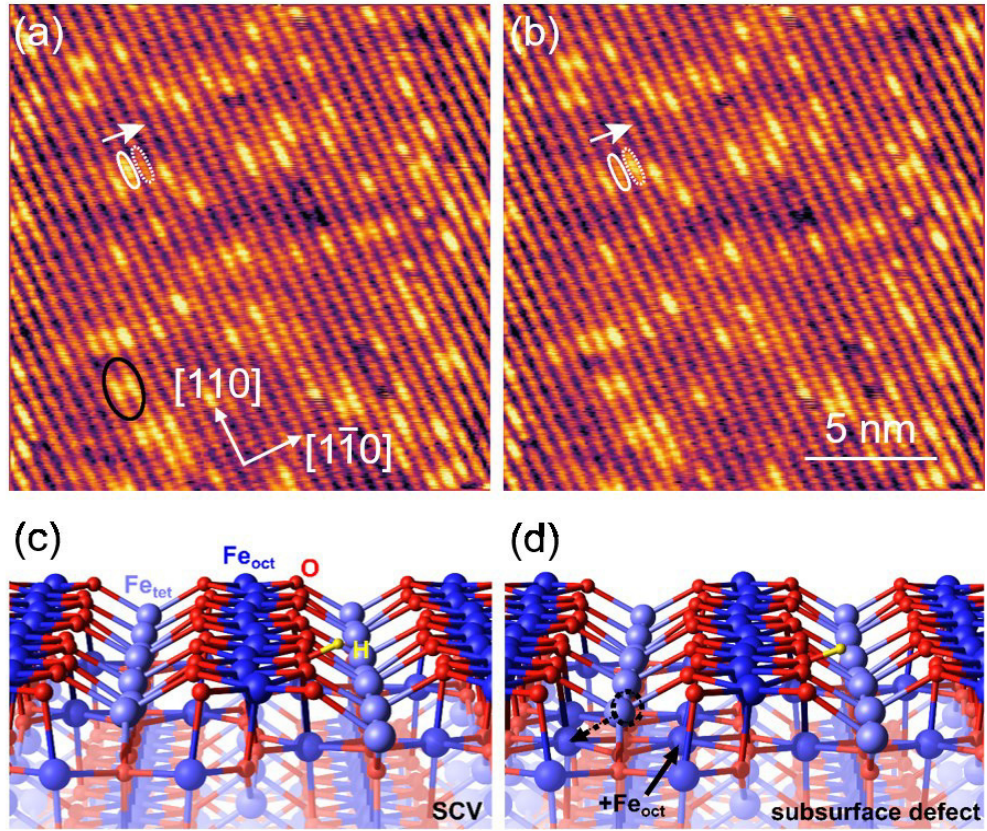

Figure 1. (a) A FastSTM frame illustrates the defects on $Fe_3O_4(001)$ discussed in the present work: a hydroxyl group resulting from a H adatom appears as an elongated protrusion (marked by a solid white oval), while an unreconstructed unit cell appears as a bright double-lobe feature (marked by a black oval). The $[1\bar{1}0]$ and $[110]$ crystal directions are marked by white arrows. (b) In the subsequent frame, the hydroxyl group has moved to its neighboring site – out of the solid white oval and into the dashed white oval. *Imaging parameters: $V_b = 1.50\ V, I_t = 1.0\ nA$.* (c, d) Structural models of the SCV-reconstructed $Fe_3O_4(001)$ surface in perspective view obtained from density functional theory (DFT) calculations. (c) shows a hydroxyl group on the pristine SCV surface and (d) a hydroxyl group close to an unreconstructed unit cell site. Note the different arrangement of the cations in the second and third atomic layer. In the bulk-like unreconstructed unit cell (d), an additional $Fe_{oct}$ is located in the third atomic layer (labelled '+$Fe_{oct}$') and another Fe atom moves from the second layer $Fe_{tet}$ site of the reconstruction into an $Fe_{oct}$ site in the third layer, as indicated by the dashed arrow.

METHODS

The Fe$_3$O$_4$(001) sample was prepared by repeated cycles of sputtering for 5 min in $5 \times 10^{-6}$ mbar Ar at 1.0 kV and subsequent annealing to 983 K in $1.0 \times 10^{-6}$ mbar O$_2$. Surface hydroxyl groups result naturally from background water dissociation during the annealing process.[15] A boron nitride heater was used to heat the sample and a type K thermocouple attached to the crystal to record the temperature. The thermocouple is referenced internally to the chamber temperature. This results in a very high relative accuracy (better than 0.1 K) while the absolute temperature value might have an offset up to 5 K, as confirmed by the order-disorder phase transition of the Fe$_3$O$_4$(001) surface at 725 K in low energy electron diffraction measurements.[25] All STM measurements were performed with an Omicron VT-AFM using etched Pt/Ir tips (Unisoku). Standard STM measurements were performed in constant current mode. A FAST module[26,27] attached to the standard Omicron controller was used to record movies at multiple frames per second. During such FastSTM measurements, the feedback was lowered to only correct for thermal drift but not follow the surface topography, and the respective scanning parameters were set before starting the FastSTM measurement. The FastSTM frame rates used for this study were either 4.0 or 19.6 frames per second (fps). Note that the sinusoidal tip movement in the fast scan direction results in a higher pixel density at the left and right edges of the movie. FastSTM data is recorded as a one-dimensional data stream and subsequently reconstructed into movies with a specially built Python script. Using this software, we level the background by FFT filtering the frequencies of the fast and slow scan directions and their overtones. We then further FFT filter the known frequencies of the setup (e.g. from turbomolecular pumps), those below the frequency of the fast scan direction, and instabilities from feedback corrections using a broad high-pass damping filter, and finally drift correct the movie. Quantitative analysis was performed by placing oval masks on one of the two locations of a hydroxyl group and integrating over their signal intensity for each frame, as demonstrated in Figure 2. The result is a telegraph noise-like graph illustrating switching between more and less bright appearances, which can be assigned to "H adatom present" and "H adatom absent", respectively. Particular care was taken to distinguish short residence times from tip changes, scratches, and the influence of other mobile defects. The residence times were then plotted as histograms, where the bin width was constrained to multiples of the respective acquisition times for one frame and a single exponential was fitted to the histogram. The mean switching rates and their respective errors were determined from the decay times fitted to all bin widths which created histograms with enough bins to properly represent the data. Typically, at least 50 events are included in a histogram at every temperature. A systematic atom tracking study on the diffusion of Si dimers showed that the electric field in a tunneling junction can influence the jump rates, but almost exclusively by the prefactor and not by the activation energy.[28] Hence, to avoid systematic errors, all FastSTM movies for H adatom switching analysis were recorded at $V_b = 1.50\ V, I_t = 1.0\ nA$.

The Vienna *ab initio* Simulation Package (VASP)[29,30] was used for all DFT calculations. The Projector Augmented Wave (PAW)[31,32] method describes the electron and ion interactions, with the plane wave basis set cut-off energy set to 550 eV. We used the following settings: a Γ-centered **k**-mesh of 5 × 5 × 1 for the bulk, $Fd\bar{3}m$, a = 8.396 Å, experimental lattice magnetite cell and (001) surface calculations (Γ-point only for 2×2 supercell). After pre-converging calculations using the Perdew-Burke-Ernzerhof (PBE)[33] functional and where dispersion effects are simply treated with D2,[34] the obtained results are further relaxed and improved using optB88-DF.[35–38] The PBE+D2[35,36] and optB88-DF van der Waals functionals were used with an effective on-site Coulomb repulsion term $U_{eff}$ = 3.61 eV[39,40] to accurately model the oxide. Calculations were performed on an asymmetric slab with 13 planes (5 fixed and 2 relaxed $Fe_{oct}O_2$ layers) and 14 Å vacuum. To avoid interaction between adsorbates, and to accurately model the experimental coverages, a (2 × 2) supercell was used (i.e. four times the (√2×√2)R45° reconstructed cell). Transition states are determined via Nudged Elastic Band[41] (NEB) calculations using the climbing-image method. The PBE+D2 functional was used for the NEB calculations, but the energies obtained at the saddle points were further improved using the optB88-DF functional keeping the ionic position fixed. Convergence is achieved when forces acting on ions become smaller than 0.04 eV/Å.

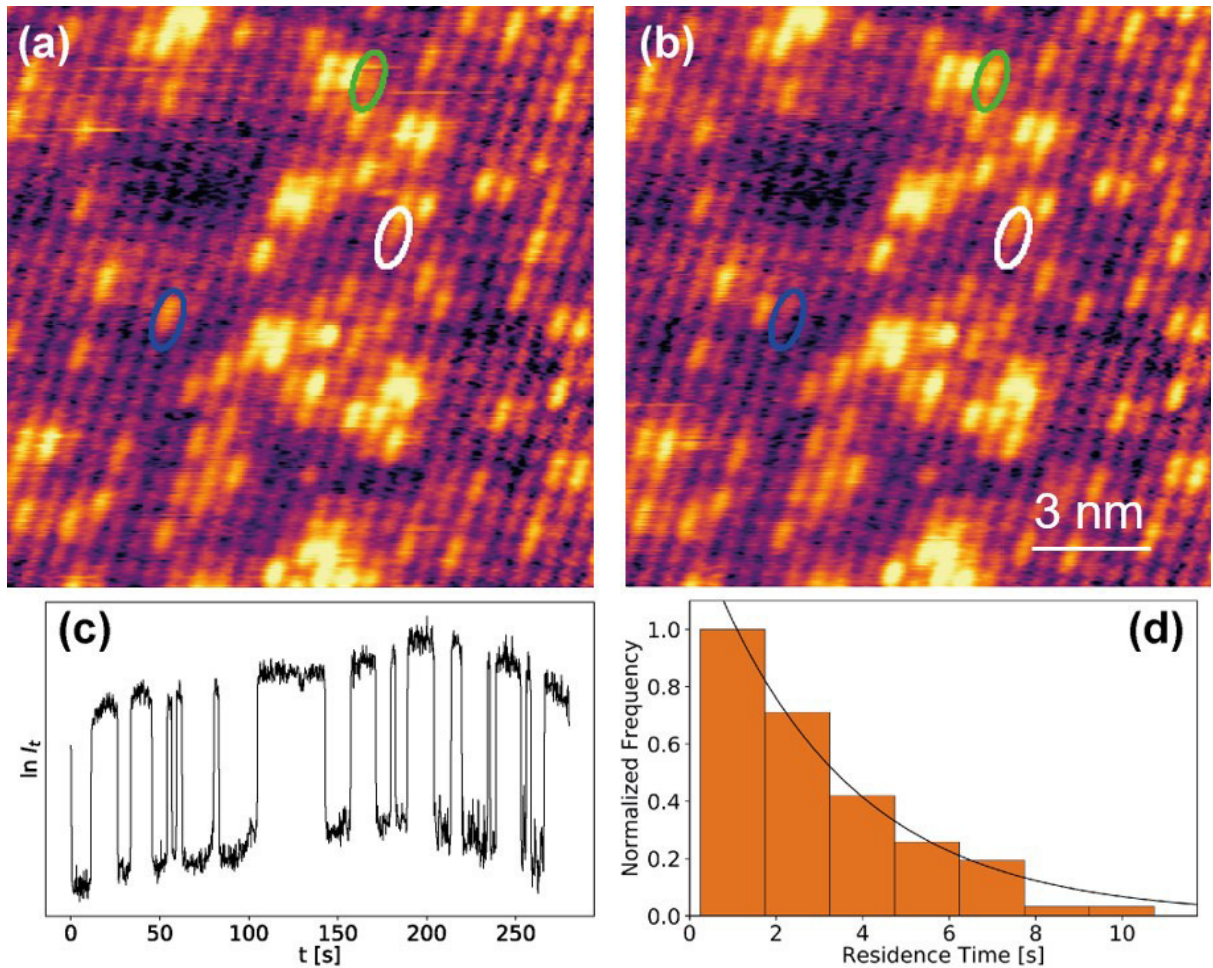

Figure 2. (a, b) Two consecutive frames from a FastSTM movie with multiple switching H adatoms. The blue mask shows a hydroxyl group on the pristine SCV surface which is located in the mask in the first frame and jumped out of it in the second frame. The green and white masks mark hydroxyl species with a defect next to them in the $[1\bar{1}0]$ and $[110]$ direction, respectively. The complete movie can be found as Supporting Movie S2. *Imaging parameters:* $V_b = 1.50\ V, I_t = 1.0\ nA, 4\ fps$. (c) Plotting the signal inside a mask as a function of time (i.e. frame), we obtain a telegraph noise-like trace for a switching H adatom. (d) Residence times extracted from all traces at a given temperature are plotted in a histogram which is then fitted by an exponential function.

RESULTS AND DISCUSSION

Making use of the increased frame rates available with FastSTM, we observed the reversible switching of H adatoms between two adjacent O atoms at room temperature and above in real time. As in previous studies at lower temperatures, we find that the H adatoms switch exclusively between the two O atoms on neighboring iron rows of the surface reconstruction (see Figure 1).[14–16,21] The process is reversible and can be monitored for many minutes to hours in temperature equilibrium. Measurements at different temperatures reveal an increase in the rate of motion with increasing temperature, implying that we are observing an activated process. Two representative movies at 331 and 349 K are shown in Supporting Movies 1 and 2, respectively. In order to determine the Arrhenius activation barrier of the switching process, we identified hydroxyl species by conventional STM imaging and then recorded FastSTM movies of these species at several temperatures. The Arrhenius plot obtained from the mean switching rate of data recorded between 336 and 382 K is shown in Figure 3. Here, only hydroxyl species which are located on a pristine SCV surface are used in the analysis, i.e. those that are located in an SCV reconstructed unit cell, as shown in Figure 1c. The rates for switching in one direction or the other are not significantly different within the statistical error of our measurements, i.e. the switching process is a symmetric one (as expected for motion between two equivalent sites). Via orthogonal distance regression, we obtain an Arrhenius activation barrier of 0.94 ± 0.07 eV and a prefactor of $1 \times 10^{13\pm1}$ Hz for hydroxyl groups on the pristine SCV surface. In order to account for entropic contributions to the activation Gibbs free energy, we also fitted the same data taking into account the transition state theory (description by Winzor and Jackson in[42]). As expected for a simple, light adsorbate like a single H atom – compared to larger, molecular adsorbate species[43] – we confirmed that the entropic contribution is negligible and the same activation energy obtained by both analysis routes.

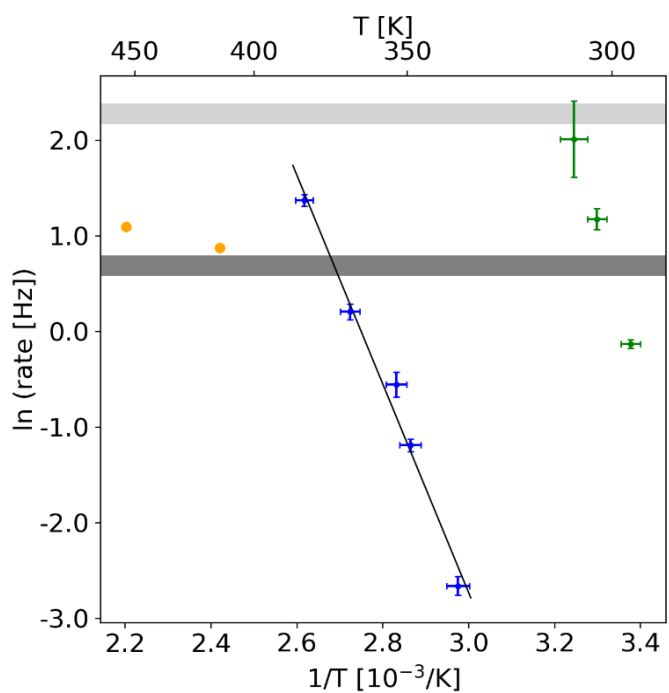

Figure 3. Arrhenius plot for H adatoms switching between two opposite O atoms. Switching rates determined from FastSTM data of species on the pristine SCV surface are shown as blue points, a linear Arrhenius fit as a black line. H adatoms next to unreconstructed unit cells in the [1$\bar{1}$0] direction exhibit a higher jump rate, shown as green points. The dark (4 fps) and light gray (19.6 fps) horizontal bars illustrate the limits of time resolution as described in the text. The orange points represent high temperature data points measured at 4 fps, demonstrating the effect of undersampling. Switching rates for hydroxyl groups on the pristine SCV surface up to 367 K were acquired at 4 fps, while those at 367 K and higher as well as those for hydroxyls next to unreconstructed unit cells were measured at 19.6 fps.

When extracting residence times from FastSTM movies, care must be taken that the frame rate is sufficient to resolve the process to be studied. As a rule of thumb, we assume that reliable data can only be obtained when the frame rate is at least twice the mean jump rate, i.e. a species is observed on average at least in two consecutive frames. Taking into account the constraints of our instrument (i.e. preamplifier bandwidth, etc.) and keeping the lateral resolution sufficient for reliable analysis, we achieved a maximum frame rate of 19.6 fps in the present work. This corresponds to a maximum time resolution of 9.8 per second, as marked by the light gray bar in Figure 3. In comparison, the more convenient frame rate of 4 fps (allowing us to scan larger areas) gives a limit of 2 per second, marked by the dark gray bar. Data points for the hydroxyl groups on the pristine SCV surface were measured with 4 fps up to 367 K and 19.6 fps at higher temperatures. At 367 K, we further confirmed that data acquisition yielded the same hydroxyl switching rate at 4 fps and at 19.6 fps. To illustrate the effect of undersampling, the switching rates obtained with 4 fps measurements at higher temperatures (413 K and 454 K) are shown as orange points in Figure 3. A clear deviation from the linear Arrhenius behavior is seen, and the obtained switching rates are close to the estimated limit marked by the dark gray bar. Indeed, these fast-moving hydroxyl groups often appear as partial protrusions in both locations in single frames. However, even at 450 K, where we cannot resolve the motion temporally, streaks between two neighboring iron rows confirm that the H adatoms only move between these two locations and do not diffuse laterally across larger distances.

Furthermore, we observed that switching rates of hydroxyl groups in the vicinity of defects differ significantly from the ones on the pristine SCV surface. On the one hand, hydroxyl species directly neighboring unreconstructed unit cells in the $[1\bar{1}0]$ direction (i.e. jumping to an iron row next to the defect, marked by the green mask in Figure 2a-b) switch more often in the temperature range considered in this study than those on the pristine SCV surface. Mean switching rates for such species are shown as green points in Figure 3. Just like for the species on the pristine SCV surface, we do not observe a statistically significant, systematic asymmetry in the switching rates towards and away from the defect. Assuming that only the path difference plays a role in the switching behavior in different local environments (substantiated by our DFT results which will be discussed below), we can estimate an Arrhenius activation barrier for OH switching next to an unreconstructed unit cell of 0.76 eV using the prefactor of $1 \times 10^{13}$ Hz which we obtained for the hydroxyl species on the pristine surface. Due to the increased rate, we are now hitting the limit of our time resolution at 308 K already, even with an acquisition rate of 19.6 fps. On the other hand, switching events of those hydroxyl groups neighboring the unreconstructed unit cell in the [110] direction (i.e. a neighbor along the row, as marked by the white mask in Figure 2a-b) are switching much less frequently. Since the events are rare, we do not observe a sufficient number of them to obtain a reliable switching rate, but we estimate the rate to be approximately 0.07 Hz at a temperature of 349 K. Finally, the influence of defects is a short-range effect. We do not observe any influence on hydroxyl groups which are located one unit cell further away from a defect.

We performed DFT calculations of hydroxyl groups on a pristine SCV surface and next to an unreconstructed unit cell to better understand the experimental results. In the case of a hydroxyl group on a pristine SCV surface, an activation barrier of 1.27 eV was obtained. Where an unreconstructed unit cell is adjacent to the hydroxyl in the [1$\bar{1}$0] direction, the activation barrier is lowered by 0.27 eV to a value of 1.00 eV. While the temperature range available in our experiment for the accelerated hydroxyl species is insufficient to obtain an Arrhenius activation barrier, we can compare the switching rate between species on the pristine SCV surface and those near a defect and relate these to the DFT results. Our calculations show that the diffusion mechanism remains the same in both cases. Assuming that the prefactor remains unchanged, the switching rate should thus be higher next to the defect. Assuming a reversible process with no entropy change, the energy difference of 0.27 eV corresponds to an approximately 100 K difference in temperature for a given rate, which is in good agreement with the experimental results shown in Figure 3. Furthermore, DFT confirms that the process is a symmetric one in both cases, with energy differences between the initial and final states of 0.08 eV and 0.06 eV for the hydroxyl group on a pristine SCV surface and that next to a defect, respectively, which are both within common fluctuations present in the calculations. The DFT results are thus consistent with the experimentally obtained jump rates and activation barriers. Analyzing the structures obtained from DFT (shown in Figure 1c-d and Figure 4), we note that the presence of the defect shortens the O–O distance in the initial state from 3.50 to 3.31 Å. In principle, this should facilitate a lower barrier, but this distance is still far too long for the H adatom to simply hop on a rigid lattice. Rather, the O–O distance must contract significantly to approximately 2.6 Å to create the transition state, where O-H bonds of 1.3 Å are simultaneously established to both the initial and target O atom. This situation is similar for both models considered here. Facilitating the transition state requires that both the O atoms and the 2$^{nd}$ nearest neighbor surface Fe atoms must move towards the H adatom, i.e. towards each other. Thus, it seems that the different barrier actually originates in the cumulative energetic cost of multiple distortions to the surrounding atoms that must occur to facilitate the hydrogen transfer. Ultimately, a lower barrier is observed in the presence of the unreconstructed unit cell because less distortion of the lattice is required to form the transition state. This is seen for example, in the Fe–O distances for the target O atom, which are approximately 0.1 Å shorter in the transition state when the defect is present than on the pristine SCV surface (compare Figure 4b and d). Interestingly, the Fe–O distances are slightly longer to the initial O atom in the presence of the defect. This relaxation seems to be compensated by the stronger interaction with the O atom labelled * in Figure 4b and d, which can move closer to the surface Fe$_{oct}$ atoms when it is coordinated to a subsurface Fe$_{oct}$ rather than an Fe$_{tet}$.

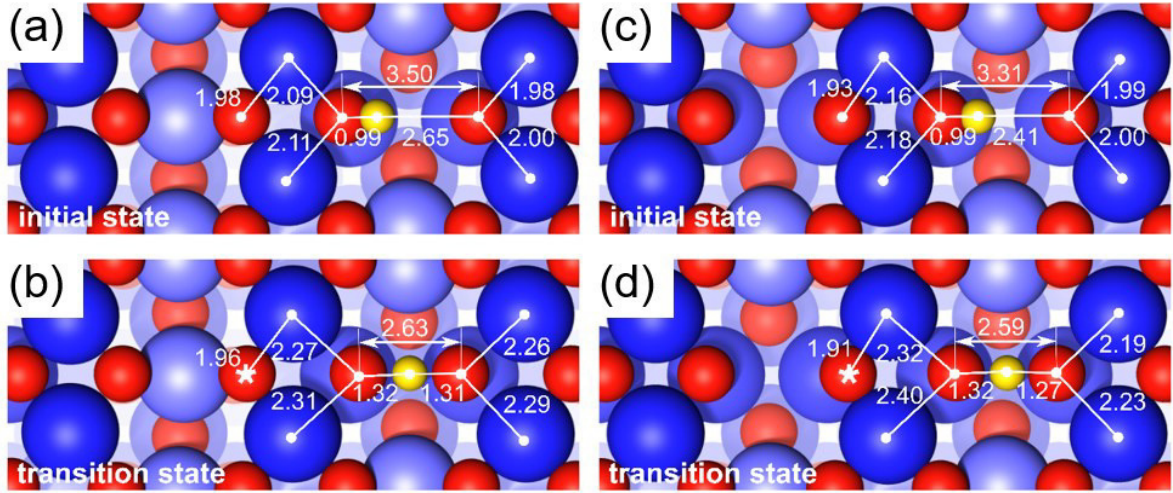

Figure 4: DFT-derived structural models illustrating the diffusion of a H adatom (a, b) on the defect-free $Fe_3O_4(001)$ surface and (c, d) neighboring an unreconstructed unit cell. Top views showing the initial and transition states for $H^+$ diffusion. The O–O distance is shorter in the initial state neighboring the defect. Nevertheless, the surrounding lattice distorts significantly in the transition state, ultimately achieving a similar O-H bond length in each case.

CONCLUSIONS

In conclusion, we have shown that the diffusion of H adatoms on $Fe_3O_4(001)$ – which occurs only between two neighboring O atoms on opposite Fe rows – is an activated process. By conducting FastSTM measurements at temperatures between 295 and 382 K, we obtained an apparent Arrhenius activation energy of $0.94 \pm 0.07$ eV for hydroxyl groups on the pristine SCV-reconstructed surface. This result is in good agreement with DFT calculations which provide an activation barrier of 1.27 eV, especially when keeping in mind that the optB88-DF van der Waals functional used here is typically over-binding[44]. Taking into account transition state theory to consider changes in entropy in the transition state, we determined that the experimentally obtained activation barrier results purely from energetics and does not contain any entropic contributions. Furthermore, we could not resolve any long-range diffusion of hydrogen up to approximately 550 K, at which point hydrogen is known to desorb from the surface.[15]

Importantly, our local measurements show that the activation barrier and thus jump rate of individual species varies significantly with their local chemical environment. This effect is generally observed for a range of different local defects such as domain boundaries or adatoms. We have studied the phenomenon in detail for hydroxyl species next to a so-called unreconstructed unit cell, i.e. a region on the surface where the interstitial $Fe_{tet}$ of the SCV reconstruction is missing, but instead two more Fe atoms are located in $Fe_{oct}$ sites in the subsurface. Our experiments showed that the switching rate of hydroxyl groups can both increase and decrease, depending on the exact location of the H adatom with respect to the defect. We explained this phenomenon by DFT calculations which show that a distortion in the Fe–O lattice brings the two neighboring O atoms closer together and thus facilitates H diffusion between them. The different energy barriers originate from the cumulative energetic cost of multiple distortions to the surrounding atoms that occur during the switching process. Less lattice distortion is thus required in the presence of defects such as a neighboring unreconstructed unit cell.

The fact that defect proximity governs the activation energy of diffusion processes highlights the importance of local techniques such as STM when it comes to investigating surface dynamics.

SUPPORTING INFORMATION

Structure of the $Fe_3O_4(001)$ surface and common defects; and diagram illustrating the energies and barriers of all processes calculated by DFT.

Supporting movie S1 shows several switching hydroxyl groups at 331 K. *Imaging parameters: $V_b = 1.50\ V, I_t = 1.0\ nA$ and 4 fps.*

Supporting movie S2 shows several switching hydroxyl groups at 349 K. *Imaging parameters:* $V_b = 1.50\ V, I_t = 1.0\ nA$ and 4 fps.

ACKNOWLEDGEMENTS

The authors would like to thank Zdenek Jakub for help with the initial sample preparation and John Ellis for useful discussions regarding the residence time distribution of binary switching events. This work was funded by the Deutsche Forschungsgemeinschaft (DFG, German Research Foundation) under research grant ES 349/1-2 and HE 3454/18-2 and under Germany's Excellence Strategy – EXC 2089/1 – 390776260, and by the EU-H2020 research and innovation programme under grant agreement no. 654360 NFFA-Europe. B.A.J.L. gratefully acknowledges a Research Fellowship from the Alexander von Humboldt Foundation, a Marie Skłodowska-Curie Individual Fellowship under grant ClusterDynamics (no. 703972) from the European Union's Horizon 2020 research and innovation program and financial support from the Young Academy of the Bavarian Academy of Sciences and Humanities. GSP and MM acknowledge funding from the Austrian Science Foundation (FWF) Start Prize Y847-N20. The computational results were achieved in part using the Vienna Scientific Cluster (VSC 3).

ORCID

| | |
|---|---|
| Barbara A. J. Lechner | 0000-0001-9974-1738 |
| Friedrich Esch | 0000-0001-7793-3341 |
| Gareth Parkinson | 0000-0003-2457-8977 |
| Ueli Heiz | 0000-0002-9403-1486 |

TOC Graphic

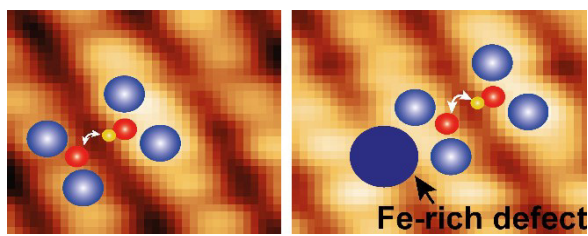

Supporting Information

for

Influence of Local Defects on the Dynamics of O-H Bond Breaking and Formation on a Magnetite Surface

*Alexander Bourgund,[1] Barbara A. J. Lechner,[1,*] Matthias Meier,[2,3] Cesare Franchini,[3] Gareth S. Parkinson,[2] Ueli Heiz[1] and Friedrich Esch[1]*

[1] Chair of Physical Chemistry, Department of Chemistry & Catalysis Research Center, Technical University of Munich, 85748 Garching, Germany

[2] Institute of Applied Physics, Technische Universität Wien, 1040 Vienna, Austria

[3] Center for Computational Materials Science, Faculty of Physics, University of Vienna, 1090 Vienna, Austria

* bajlechner@tum.de

### a) Structure of the Fe₃O₄(001) surface and common defects

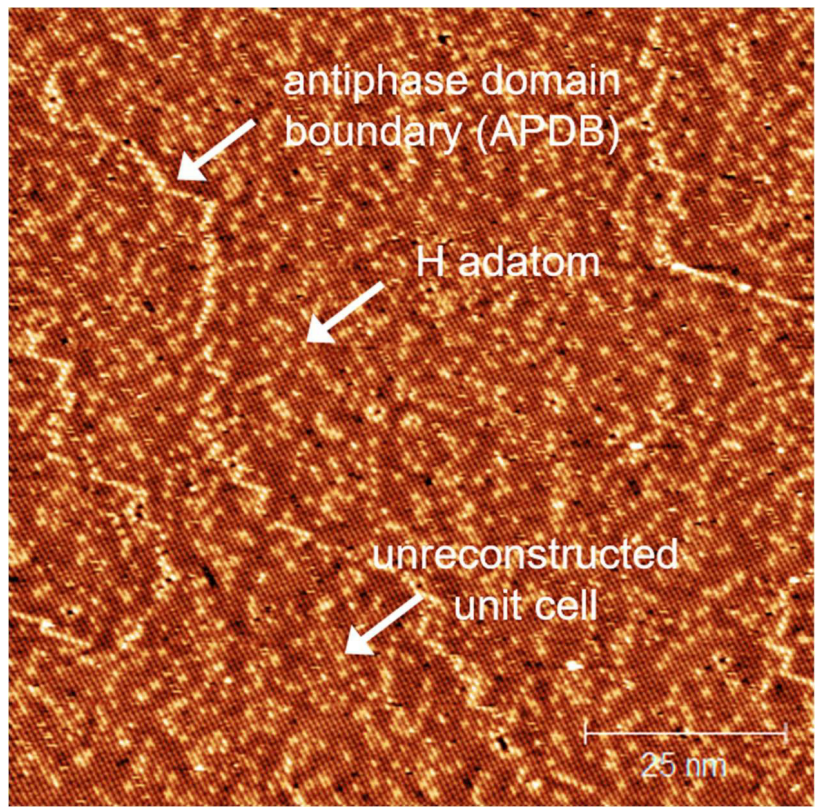

Figure S1: A large-scale STM image shows a typical preparation of the $Fe_3O_4$(001) surface, including the most common defects (marked by white arrows). The undulating Fe rows are clearly visible through electronic contrast (the O atoms of the surface are not observed by STM), oriented here in the direction from the top left of the image to the bottom center. We find a monophase termination of one large terrace, with several domains of the (√2×√2)R45° reconstruction, separated by antiphase domain boundaries which zigzag across the rows in a 45° orientation. *Imaging parameters:* $V_b = 1.50\ V, I_t = 0.3\ nA$.

**b) Diagram illustrating the energies and barriers of all processes calculated by DFT**

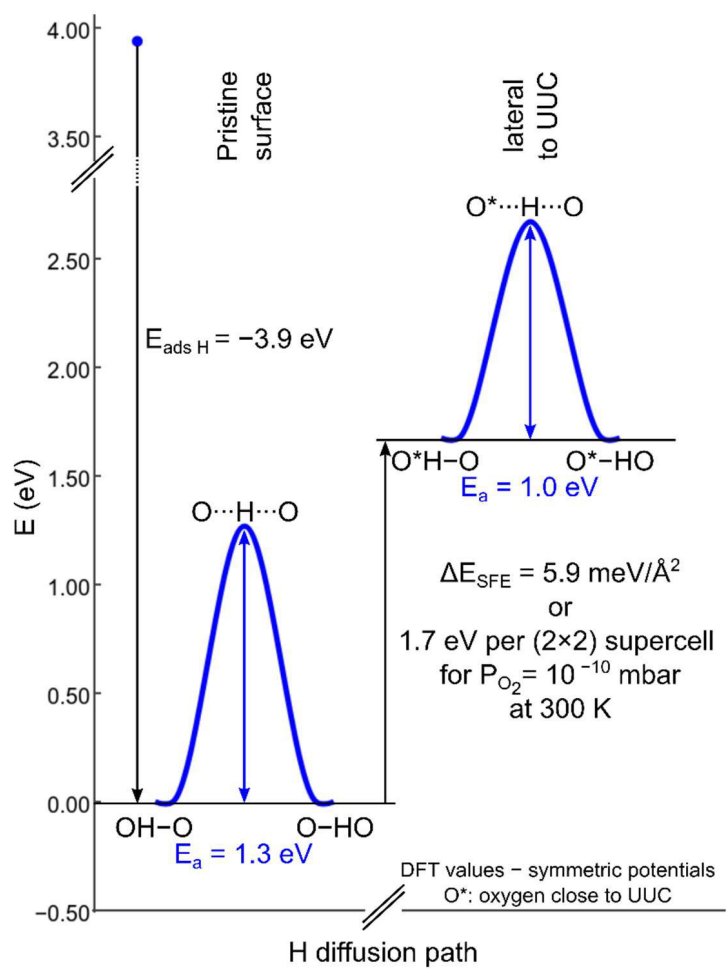

Figure S2: Energy diagram summarizing the energies and barriers from DFT calculations. The left-hand side shows the energy barrier for the switching process of a hydroxyl group on the pristine surface, while the case of a switching hydroxyl group next to an unreconstructed unit cell (UUC) defect is shown on the right-hand side. The adsorption energy of H is given with respect to H (rather than H$_2$) in the gas phase as $E_{ads} = E_{slab\ with\ H} - E_{clean\ slab} - E_{H\ gas}$. The difference in surface free energy (SFE) between the two cases is given as $\Delta E_{SFE}$.[1,2] The SFE is a function of the chemical potential of O and H, but since only a relative difference is required, the chemical potential of H can be set constant as the amount of H is the same in both cases. The chemical potential of O is taken for a pressure of O$_2$ of $p_{O_2} = 10^{-10}$ mbar at 300 K, corresponding to experiments under ultra-high vacuum conditions.[3] To estimate

roughly how much the formation of a defect costs energetically, we multiply the SFE difference of $\Delta E_{SFE} = 5.9$ meV/Å² with the area of a 2 × 2 supercell, i.e. assuming one defect every four unit cells (an upper limit for defect density which can realistically be expected to occur). As expected, we find that the defect is unfavorable compared to the SCV reconstruction, irrespective of the presence of an H adatom.